\documentstyle[prd,tighten,aps]{revtex}
\begin{document}
\draft

%begin wide text
\twocolumn[\hsize\textwidth\columnwidth\hsize\csname
@twocolumnfalse\endcsname
\renewcommand{\theequation}{\thesection . \arabic{equation} }
\title{\bf Observable effects from spacetime tunneling}

\author{Pedro F. Gonz\'alez-D\'{\i}az}
\address{Centro de F\'{\i}sica ``Miguel Catal\'an'',
Instituto de Matem\'aticas y F\'{\i}sica Fundamental,\\
Consejo Superior de Investigaciones Cient\'{\i}ficas,
Serrano 121, 28006 Madrid (SPAIN)}
\date{August 12, 1997}

\maketitle

\begin{abstract}

Assuming that spacetime tunnels -wormholes and ringholes-
naturally exist in the universe, we investigate the
conditions making them embeddible in Friedmann space,
and the possible
observable effects of these tunnels, including: lensing and
frequency-shifting of emitting sources, discontinuous change of background
temperature, broadening
and intensity enhancement of spectral lines, so as a dramatic
increase of the luminosity of any object at the tunnel's throat.

\end{abstract}


%end wide text
\vskip2pc]

\renewcommand{\theequation}{\arabic{section}.\arabic{equation}}

\section{\bf Introduction}

Solutions to Einstein equations corresponding to spacetimes with closed
timelike curves (CTCs) have stirred the relativistics,
following developments by Lanczos [1], van Stockum [2], G\"odel [3],
Misner [4] and, more recently, Morris and Thorne [5], Gott [6]
and Jensen and Soleng [7]
The reason for the excitement and subsequent general
disbelief resided much in that,
being consistent solutions to Einstein equations for very special
kinds of matter, the proposed spacetimes allow for the possibility
of time travel and, hence for potential violations of causality [8].
Here we look at the idea that, rather than being interpreted as
constructs to be eventually built up from future highly
developed technology, spacetime tunnels generating
CTCs may spontaneously exist in some regions
of our universe, and give rise to observable effects which could
be detected even with present technology.

Thus, we consider CTCs generated along spacetime tunnels whose mouths
embed in distant regions with the Friedmann geometry of the overall
universe. The necessary condition for such tunnels to occur in
a given region is that, in that region there is a certain proportion
of matter with negative energy [8,9]. Two tunnel topologies have been
considered so far. That of a two-sphere which gives rise to
wormholes [5,10], and that of a two-torus which is associated with the
so-called ringholes [11]. Both tunneling types are traversable and
convertible into timemachines generating CTCs by simply letting
one of the hole's mouths to move toward the other [5,10,11], but
whereas the energy density is everywhere negative near the throat
of a wormhole, it still becomes positive for values of the angle
$\varphi_2$ (defining the position on the surface circles determined
by the torus sections) such that $2\pi -\varphi_h >\varphi_2 >
\varphi_h$, with $\varphi_h=\arccos\frac{b}{a}$, where $a$
and $b$ are the radius of the circumference generated by the
circular axis of the torus and that of a torus section, respectively,
at the ringhole's throat [11].

The purpose of the present work is to investigate under what
conditions can a tunnel be embedded in a cosmological spacetime,
and explore the effects
that the inner properties of spacetime tunnels may have on the
observable characteristics of astronomical
objects placed beyond, or passing through these tunnels,
relative to an observer whose line of sight to the object traverses
or does not traverse
the given tunnel. Most of the emphasis will be placed on ringholes,
but the results will be always compared
with those expected from wormholes.

\section{\bf Embedding a tunnel in Friedmann space}

Assuming that the amount of negative mass equals that of positive
mass, the gravitational field that represents tunneling through
a traversable ringhole can be described by a spacetime metric [11]:
\begin{equation}
ds^2=-dt^2+(\frac{n}{r})^2dl^2+m^2d\varphi_1^2+b^2d\varphi_2^2,
\end{equation}
where $-\infty < t <\infty$, $l$ is the proper radial distance
of each transversal section of the ringhole's torus,
$-\infty < l <\infty$, $\varphi_1$ and $\varphi_2$ are
as given for the torus metric [11], and
\[m=a-b\cos\varphi_2,\;\; n=b-a\cos\varphi_2,\]
\begin{equation}
r=\sqrt{a^2+b^2-2ab\cos\varphi_2}.
\end{equation}

There are two $\varphi_2$-angular horizons. They occur at
$\varphi_2=\varphi_h=\arccos\frac{b}{a}$ and $2\pi-\varphi_h$.
For $-\varphi_h<\varphi_2 < \varphi_h$, the ringhole would behave like
a converging lens near the throat, and for
$2\pi-\varphi_h >\varphi_2 > \varphi_h$,
the ringhole would behave like the wormhole; i.e.
it acts as a diverging lens near the throat. This property may be
crucial to identify the existence of these holes in the universe
and distinguish between them.

Metric (2.1) can easily be converted into that of a spherical
wormhole by applying the coordinate change [11]
$a\rightarrow 0$, $\varphi_2\rightarrow\theta+\frac{\pi}{2}$,
$\varphi_1\rightarrow\phi$, and allowing a given purely
negative mass $M$ to induce a general factor $e^{2\Phi(r,M)}$
in the $g_{tt}$ components of the metric tensor.

For a section of constant $\varphi_1$ and
$\varphi_2$, the ringhole metric (2.1) can be written as
\begin{equation}
ds^2=b^2(-d\eta^2+d\chi^2),
\end{equation}
where
\begin{equation}
d\eta=\frac{dt}{b},\;\;\; d\chi=\frac{ndl}{rb}.
\end{equation}
In the approximation in which $b\gg b_0$ (where $b_0$ is the value
of $b$ at the throat), $l\sim b$ [11], one can integrate the
second of expressions (2.4) to
\[\chi=\chi_0\]
\begin{equation}
+\ln\left[\left(\frac{r}{a}+\frac{b}{a}-\cos\varphi_2\right)
\left(\frac{a}{b}+\frac{r}{b}-\cos\varphi_2\right)^{\cos\varphi_2}\right].
\end{equation}
When a null geodesic is considered, $ds=0$ and we obtain
\[b'=\frac{db}{d\eta}=b\frac{db}{dt}=\frac{rl}{n}\]
\[d\eta=\pm d\chi ,\]
or integrating
\begin{equation}
\chi=\pm\eta+\chi_0,
\end{equation}
and hence, following the same line of reasoning as in e.g. Ref. [12],
we obtain that the frequency of the light rays is either red-shifted
while traveling from mouth to throat by an amount
\begin{equation}
z=\frac{\omega-\omega_0}{\omega_0}
=-\chi\frac{b'}{b}=-\chi\frac{rl}{nb},
\end{equation}
or blue-shifted by exactly the same amount if the light rays travel
from throat to mouth. Therefore, along complete passage through a
static ringhole, the frequency of a light ray preserves its initial
value. However, the embedding of the ringhole mouths in two
respectively distant regions of a Friedmann spacetime with scale
factor $R(\eta_{F})$, conformal time
$\eta_{F}=\int\frac{dt_F}{R}$ and distance from
the origin $R\chi_{F}$, ensures that such two distant regions are
red-shifted relative to one another by the known cosmological
amount predicted from
$\omega=\omega_0 R(\eta_{F}-\chi_{F})/R(\eta_{F})$ [12],
both for static ring- and worm-holes. It follows that these
tunnels cannot embed in the background Friedmann spacetime,
unless for the case that the tunnels be allowed to be no longer
static, with one
of the hole's mouths moving with respect to
the other mouth with a nonzero velocity,
$V$, such that the Doppler shift resulting from this motion
would equal the usual cosmological red-shift:
\begin{equation}
z=-\frac{HD}{c}=-\frac{V}{c},
\end{equation}
with $H$ the Hubble's constant and $D$ the distance, both for
worm- and ring-hole. In the latter case, the initial
distance between mouths can be calculated to be:
\begin{equation}
L_r=2\int\frac{nb}{rl}db=2a,
\end{equation}
where we have discarded the integration constant,
so that (2.8) can finally be re-written as
\begin{equation}
z=-\frac{HD}{c}-\frac{2H_r a}{c},
\end{equation}
with $H_r=\frac{b'}{b^{2}}$.

The metric of a ringhole whose mouths move with velocity $V$
relative to one another has the form [11]
\begin{equation}
ds^2=-\left[1+glF(l)\sin\varphi_2\right]^2 dt^2
+dl^2+m^2d\varphi_1^2+b^2d\varphi_2^2,
\end{equation}
where $g=\gamma^2(dV/dt)$, $\gamma=1/\sqrt{1-V^2}$, $l=\sqrt{b^2-b_0^2}$,
with $b_0$ the value of $b$ at the throat, and $F(l)$ is a form
factor that, if we assumes mouth A to be moving, vanishes in
the half of the ringhole with the other mouth, and rises
monotonously from 0 to 1 as one moves from the throat to
mouth A. For this ringhole to be embedded in Friedmann spacetime,
the velocity $V$ must be restricted to be
\begin{equation}
V=HD=\frac{\dot{R}}{R}D,
\end{equation}
with the overhead dot meaning derivative respect to the
Friedmann time $t_F$.
In this case, the embedding of sections in the ringhole space
with constant angles
$\varphi_1$ and $\varphi_2$ in sections of Friedmann space with
constant angles $\theta$
and $\phi$, implies
\begin{equation}
dt_F=(1+glF(l)\sin\varphi_2^{(0)})dt,
\end{equation}
with $t_F$ and $t$ the times for, respectively, Friedmann
and ringhole spacetimes.

Let us evaluate $g$ at given constant values of $\varphi_2^{(0)}$
and $D$. We have
\[g=\gamma^2\frac{dV}{dt}=
\gamma^2 D\left(\frac{dH}{dt_F}\right)(1+glF(l)\sin\varphi_2^{(0)}),\]
where (2.13) has been used. From this we obtain
\begin{equation}
g=\frac{\gamma^2 DH^2 (q-1)}{1-\gamma^2 DH^2 (q-1)lF(l)\sin\varphi_2^{(0)}},
\end{equation}
where $q=-\frac{\ddot{R}R}{\dot{R}^2}$ is the deceleration
parameter at the time when the ringhole is formed.
Hence, we obtain as the metric of a ringhole embeddible in Friedmann
space:
\[ds^2=-\left[1-\frac{DH^2(q-1)lF(l)\sin\varphi_2^{(0)}}{(1-H^2 D^2)^2}\right]^{-2} dt^2\]
\begin{equation}
+dl^2+m^2 d\varphi_1^2+b^2 d\varphi_2^2.
\end{equation}

The corresponding embeddible metric for wormholes
can again be obtained from
this by using the transformation [11] $a\rightarrow 0$, $\varphi_2
\rightarrow\theta+\frac{\pi}{2}$, $\varphi_1\rightarrow\phi$,
and introducing a factor $e^{2\Phi}$ in the $g_{tt}$ component
of the metric tensor.
On the other hand, a ringhole embeddible in Friedmann space will
develop closed timelike curves and hence convert into timemachine
at sufficiently late times, provided that the time shift
induced by relative motion between mouths exceeds the distance
between mouths [9], i.e.
\begin{equation}
t_F>\left|\frac{1-H^2D^2}{(q-1)HlF(l)H_r\sin\varphi_2^{(0)}}\right|,
\end{equation}
where Eqns. (2.10), (2.13) and (2.14) have been used. Thus, for a given
time $t_F$, and provided that there exists a certain proportion
of matter with negative energy, a ringhole timemachine can be
spontaneously created for the set of parameters $l$, $F(l)$,
$H_r$ and $\varphi_2^{(0)}$ which satisfy (2.16), so that
any astronomical object going through the ringhole should
traverse CTCs in the inner nonchronal region of the ringhole
to travel back- or for-ward in time, depending on whether it enters
the ringhole by its moving or stationary mouth [11].

\section{\bf Observable effects}
\setcounter{equation}{0}

\subsection{\it Lensing}

Let us assume that at one of the ringhole's mouths
there is a source of radiation which, since the hole is traversable,
can be detected by an observer placed at the other mouth. Some of
the rays emitted by the source will cross the ringhole's throat on one of
the angular horizons, $\varphi_h$ say,
where they will not undergo any deflection [11].
Other rays will however cross the throat along values of the angle
$\varphi_2$ such that $2\pi-\varphi_h >\varphi_2 > \varphi_h$ and,
therefore, will
be deflected towards the angular horizon by gravitational repulsion
of the negative energy placed on this horizon side. One should
then expect that the observer will detect a double image of the
source, such as it is described in Fig. 1. Assuming that the
total amount of negative mass is $M<0$, we then have
\begin{equation}
\delta=2\alpha=\frac{4G|M|}{b(\pi-\varphi_2)c^2}.
\end{equation}
Besides, for small $\alpha$ it can be obtained
\begin{equation}
\alpha=\frac{\varphi_2-\varphi_h}{L_N}b=\frac{\varphi_2-\varphi_h}{d}2b.
\end{equation}
From (3.1) and (3.2),
\begin{equation}
\epsilon^2-(\epsilon_h+\epsilon_{\pi})+(1+\epsilon_h\epsilon_{\pi})=0,
\end{equation}
where
\[\epsilon=\frac{b\varphi_2}{\sqrt{dR_s}},\;\;
\epsilon_\pi=\frac{\pi b}{\sqrt{dR_s}},\;\;
\epsilon_h=\frac{b\varphi_h}{\sqrt{dR_s}},\]
with $R_s=2G|M|/c^2$.
The solutions to the quadratic (3.3) are
\begin{equation}
\epsilon_{\pm}=\frac{\epsilon_h+\epsilon_{\pi}\pm\sqrt{(\epsilon_h-\epsilon_{\pi})^2-4}}{2}.
\end{equation}
It follows that: (i) if $\epsilon_{\pi}-\epsilon_h > 2$ there will be
two rays deflected to the observer, (ii) if $\epsilon_{\pi}-\epsilon_h < 2$
deflection will fully block all rays from reaching the observer,
and finally (iii) if $\epsilon_{\pi}-\epsilon_h = 2$ there will be
an umbra region surrounded by a caustic where rays accummulate to
produce a brilliant region of highly enhanced intensity [13]. Double
imaging in ringholes would tend to occur for generally small
mass $|M|$ and mouth separation $d$, and generally large radius
of the throat. The effect would also be enhanced when $b$
approaches $a$.

In the above analysis we have assumed that direct rays linking
the source to the observer pass on the angular horizon at or
near the ringhole's throat. Moreover, it can be shown
that if the velocity
of the particles coming to the detector is $v$, then the
observer will detect a second image of the source with a
shifted frequency
\begin{equation}
\frac{\delta\omega}{\omega}=\frac{\gamma vG|M|}{b(\pi-\varphi_2)c^2},
\end{equation}
where $\gamma$ is the relativistic factor for velocity $v$, such
as it happens in cosmic strings [14].

The lensing and frequency-shifting effects could not occur
in wormholes, where one must assume the negative energy to be
uniformly or radially distributed in the region surrounding the throat.
However, they should happen in ringholes even when the mouths
are at rest relative to one another. The latter is the case
which is actually assumed in the above calculation and
Fig. 1. For an embeddible ringhole, the calculation goes along
similar steps, with $\delta=\alpha+\beta$ ($\beta\neq\alpha$)
but, since both $\alpha$ and $\beta$ should be very small, the
final result cannot appreciably differ from (3.4).

\subsection{\it Discontinuous change of temperature}

If the ringhole is allowed to have its two mouths in relative
motion at a speed $V$,
setting two objects initially at rest
on one side of the throat, but near to it, one on the angular
horizon and the other at an angle
$2\pi-\varphi_h >\varphi_2 > \varphi_h$, then
when the mouth on the other side starts moving, there will appear
a transverse velocity component for the object
at $2\pi-\varphi_h >\varphi_2 > \varphi_h$
toward the object on the angular horizon due to the diverging
effect of negative mass. This velocity component is given by
\[u=\gamma_{V}V\delta,\]
with $\gamma_{V}$ the relativistic factor for velocity $V$ and
$\delta$ as given by (3.1).
Letting the object on $2\pi-\varphi_h >\varphi_2 > \varphi_h$
be a source of radiation and the object on the angular horizon $\varphi_h$
an observer, we deduce that the latter must detect a discontinuous
change of radiation frequency due to Doppler shift.

In the cosmological context, the ringhole will be backlit by
a uniform black body radiation background, so that the Doppler
shift would result in a discontinuous change of temperature,
\begin{equation}
\frac{\delta T}{T}=\frac{\gamma_{V}vG|M|}{b(\pi-\varphi_2)c^2},
\end{equation}
originated from the motion of one mouth relative to the other that
can make the source and observer traverse the throat of
the ringhole with moving mouths, in a similar effect to that is also
induced by cosmic strings [14].

This discontinuous change of temperature would not be expected to
happen in wormholes where the throat region is filled with negative
energy only and therefore there is no angular horizons.

\subsection{\it Line broadening}

On the other hand, assuming the usual Maxwellian
velocity distribution, the relative probability for atoms with
resonant emission frequency $\omega_0$ [15], entering one
mouth of a ringhole with its mouths in relative motion at
velocity $V$, to have the $Z$-component (we assume
$Z$ to be the main axis of the ringhole) of their velocity with
values between $v_Z$ and $v_{Z}+dv_Z$ will be given by
\begin{equation}
\exp\left[-\frac{m_A(\pi-\varphi_2)bc^3}{2k_B \gamma_{V}VR_s \omega_0^2 T_{MW}}(\omega-\omega_0)^2\right]\frac{cd\omega}{\omega_0},
\end{equation}
where Eqn. (2.8) has been used, $m_A$ is the mass of the atom and
$T_{MW}$ is the temperature of the microwave background. So, the
lines of the spectrum will have their maximum at $w_0$ and a full
Doppler-broadened width at half their maximum height
\begin{equation}
\frac{4\omega_0 k_B \gamma_{V}VR_s T_{MW}\ln 2}{m_A b(\pi-\varphi_2)c^3}.
\end{equation}
A large broadening would then be expected for spectral lines from
atoms emitting through a ringhole embedded in Friedmann space,
from values of
$\varphi_2$ close to $\pi$, but not through any kind of wormholes.

\subsection{\it Spectral intensity and luminosity}

Finally, let us consider the effect that
spacetime tunnels may have in the spectral intensity of distant
objects. If a number $N$ of emitting two-level atoms cross a
ringhole, they will pass through some regions with negative
energy and some regions with positive energy near the throat.
In regions with negative energy chaotically assambled, the
temperature will be negative and give rise to population
inversion in the energy levels of the atoms, in a direct effect
which differs from the pumping by a three-level systems that
most simply characterizes lasers [15].
Instead, here the population inversion
is originated by the fact that in the Boltzmann's law
\[N_1=N_2\exp\left(\frac{\hbar\omega}{k_B T}\right),\]
(where $N_1$ and $N_2$ are the population of, respectively,
the first and second atomic levels, assumed to be nondegenerate)
negative values of $T$ necessarily imply $N_2 > N_1$. In this
case, the beam intensity $I$
\begin{equation}
\frac{\partial I}{\partial Z}
=+(N_2-N_1)F(\omega)\left(\frac{\hbar\omega B}{cn\Gamma}\right)I,
\end{equation}
(where [15] $F(\omega)d\omega$ is the fraction of transitions in
which the photon frequency lies in a small range $d\omega$
about frequency $\omega$, $B$ is the Einstein absorption
coefficient, $n$ is the refractive index and $\Gamma$ is the
total volume of the considered region) will grow with distance
along the ringhole, through the atomic gas. Meanwhile, in the
region with positive energy, $N_1 > N_2$, and hence $I$ must
decrease with distance along direction $Z$, through the atomic gas.
Along a distance $\bigtriangleup Z\simeq\sqrt{2}b_0$ on the
line of sight, one would then expect a large increase of the
intensity of the light coming from
the compactified regions of galaxies which are confined
to a width of the order $b_0$, when the given galaxies cross a
ringhole's throat.

On the other hand, as far as we are dealing with a ringhole
which is embeddible in some region of the universe,
the entering of the galaxy in one mouth implies
its simultaneous exit at the other mouth [11], so that the overall
observable effect will be that of a very brilliant compact region of
size $\sim b_0$, embedded in the larger galaxy. This effect
would be expected for all considered tunnels, though it
became most apparent in the case of ringholes because of
the existence of angular horizons which make quite neater
the separation of the brilliant region from a darker background.

\section{\bf Perspective effects}

All the effects discussed so far refer to the particular
case where the observer's line of sight to the object passes
through the traversable spacetime tunnels. However, direct
access by sight to the region surrounding the ringhole's throat
by an observer sustends a given, more or less deformed, cone
of direct observation, generated by revolution around an axis
parallel to the ringhole axis at the center of a transversal
section of the torus, with an angle whose vertex is at
the throat. The maximum value of that angle is expected to be (Fig. 2)
\[\vartheta\approx\arctan\left(\frac{L_{r}}{2a}\right)=45^{\circ},\]
where (2.9) has been used.

Observers whose line of sight to the throat lies outside that
cone will have only a projected evidence of the object's passing
through the tunnel throat and, in the case of a static ringhole or
in the more realistic case of
a ringhole with its mouths set in relative motion with nonzero
velocity, they would detect the object reaching the two
mouths, as well as some secondary effects induced from the
primary effects occurring around the throat (strong broad
emission lines, very high luminosities induced by the smallness
of the throat, etc), but not these primary effects themselves.
Among the possible secondary effects one may include the
existence of regions filled with ionized gas which would
be observable whatever angle between the line of sight
and the ringhole axis may be formed. These regions should
be originated from the high-intensity beam of ionizing photons
possessing the whole collection of electronic transition
frequencies of the atoms and molecules whose electronic
energy levels were inversely populated in the throat region.

It is rather a curious fact that the most remarkable features
of the emerging overall picture (see Fig. 2)
can, at least qualitatively, be accommodated to the current
unified models of galaxies with active nuclei, class 1 and
class 2 Seyferts and quasars [16].
These models attribute the differences among these objects
to the presence of a dusty torus of dense molecular gas
surrounding a black hole. The dusty torus serves as well
to give rise to a cone of direct observation with angle
of about 45$^{\circ}$. Just like for ringholes, all
differences can then be attributed to merely being
observed by lines of sight which lie inside or outside this cone.

\acknowledgements

\noindent For useful comments, the author thanks M. Moles of IMAFF. This
research was supported by DGICYT under Research Projects No.
PB94-0107 and No. PB93-0139.

\vspace{.6cm}

\noindent {\bf Legend for Figures}

\vspace{.5cm}

\noindent {\bf Fig. 1}. Geometry for gravitational lensing by the
negative mass placed at the throat of a ringhole. Off-axis light
rays from stellar source $S$ are deflected to detector $D$ by the
gravitational repulsion of negative mass $M$. For embeddible
ringholes the two angles denoted $\alpha$ would become
slightly different.

\vspace{.5cm}

\noindent {\bf Fig. 2}. Schematic picture of a galaxy being
observed through a ringhole tunnel with moving mouths. The
throat can only be directly accessed if the line of sight lies
inside a direct observation cone, $\vartheta < 45^{\circ}$.
For lines of sight outside that cone, $\vartheta > 45^{\circ}$,
only the regions placed near the ringhole's mouths and some
secondary effects induced on the throat could be observed.
On the figure, HIBLR: hight-intensity broad-line region ($\circ$)
and LINLR: low-intensity narrow-line region ($\bullet$).
The term "Friedmann connection", also on this figure,
merely denotes the conventional spacetime that connects
the two mouths of the ringhole by going outside the tunnel.

\end{document}